\documentclass[aps,superscriptaddress,nofootinbib,preprint,tightenlines,showpacs,eqsecnum]{revtex4}
\usepackage{hyperref}
\usepackage{epsfig,rotating}
\usepackage{amsmath,amssymb}
\usepackage{dsfont}

\newcommand{\vp}{\vec{p}}

\newcommand{\pslash}{\not\!{p}}

\newcommand{\be}{\begin{equation}}
\newcommand{\ee}{\end{equation}}
\newcommand{\bea}{\begin{eqnarray}}
\newcommand{\eea}{\end{eqnarray}}

\begin{document}

\title{Unsterile-Active Neutrino Mixing: Consequences on Radiative Decay and Bounds from the X-ray Background}
\author{D. Boyanovsky}
\email{boyan@pitt.edu} \affiliation{Department of Physics and
Astronomy, University of Pittsburgh, Pittsburgh, PA 15260}
\author{R. Holman} \email{rh4a@andrew.cmu.edu}
\affiliation{Department of Physics, Carnegie Mellon University,
Pittsburgh, PA 15213}
\author{Jimmy A. Hutasoit}\email{jhutasoi@andrew.cmu.edu}  \affiliation{Department of Physics, Carnegie Mellon University, Pittsburgh, PA 15213}

\date{\today}

\begin{abstract}
We consider a sterile neutrino to be an unparticle, namely an \emph{unsterile neutrino}, with anomalous dimension $\eta$ and study its mixing with a canonical active neutrino via a see-saw mass matrix. We show that there is \emph{no unitary} transformation that diagonalizes the mixed propagator and a field redefinition is required. The propagating or ``mass'' states correspond to an unsterile-like and active-like mode. The unsterile mode features a complex pole or resonance for $0 \leq \eta < 1/3$ with an ``invisible width'' which is the result of the decay of the unsterile mode into the active mode and the massless particles of the hidden conformal sector. For $\eta \geq 1/3$, the complex pole disappears, merging with the unparticle threshold. The active mode is described by a stable pole, but ``inherits'' a non-vanishing spectral density above the unparticle threshold as a consequence of the mixing. We find that the \emph{radiative} decay width of the unsterile neutrino into the active neutrino (and a photon) via charged current loops, is \emph{suppressed} by a factor $\sim  \Big[2 \sin^2(\theta_0) \, \frac{M^2}{\Lambda^2}\Big]^\frac{\eta}{1-\eta}$, where $\theta_0$ is the mixing angle for $\eta=0$,  $M$ is approximately the mass of the unsterile neutrino and $\Lambda \gg M$ is the unparticle-scale. The suppression of the radiative (visible) decay width of the sterile neutrino weakens   the bound on the mass and mixing angle
from the X-ray or soft gamma-ray background.
\end{abstract}


\maketitle

\section{Introduction}\label{sec:intro}

Neutrino masses and oscillations are now an established phenomenon
and an undisputable evidence of physics beyond the standard model.
Although the origin  and scale of masses remains a challenging
question, the see-saw mechanism provides a compelling explanation of
small active neutrino masses as the result of ratios of widely
different scales \cite{seesaw}. Among extensions of the standard
model, the addition of sterile neutrinos, namely SU(2) singlets
with a mass in the few keV range, acquires particular
importance as a potential warm dark matter
candidate \cite{dw,colombi,este,shapo,kusenko,kusepetra,petra,coldmatter,micha,boysnudm,kuseobs,hector,wu}
and could provide possible solutions to a host of astrophysical
problems \cite{kusenko}. The radiative decay \cite{pal} of a
sterile-like neutrino mass eigenstate into an active-like mass
eigenstate and a photon   leads to a decay line that could be
observable in the X-ray or soft gamma ray background \cite{dolgov}.
The non-observation of this line provides a constraint on the mass
and mixing angle of sterile-like
neutrinos \cite{dolgov,sorensen,aba,kuseobs,micha,hansen}.

More recently sterile neutrinos with mass in the $\mathrm{GeV}$ range have been proposed as
explanations of two seemingly unrelated and unexpected phenomena: an excess of air shower events
at the SHALON gamma ray telescope, in a configuration where the expected number of events
is negligible \cite{shalon}, and as a potential explanation of the MiniBoone anomaly \cite{miniboone}, namely
the prominent peak of electron-neutrino events  above background for $300 \, \mathrm{MeV} \leq E_\nu \leq 475
\, \mathrm{MeV}$. Sterile neutrinos in a three-active, two-sterile ($3+2$) oscillation scheme were proposed in Ref. \cite{maltoni} as a possible
explanation of the MiniBooNE anomaly,  and  an alternative
explanation invoking the radiative decay of a heavy sterile neutrino with a small magnetic moment was proposed in Ref. \cite{gninenko}.

In this article we study the possibility that sterile neutrinos
are a manifestation of \emph{unparticles}, which then mix  with the active
neutrinos via a see-saw-type mass matrix.

In a recent series of articles, Georgi \cite{georgi} suggested an extension of the standard model
 in which particles couple to a conformal sector with a
non-trivial infrared fixed point acquiring (large or non-perturbative) anomalous dimensions with potentially
relevant consequences, some of which may be tested at the Large Hadron Collider (LHC)  \cite{cheung,raja,quiros}.  Early work by Banks
and Zaks  \cite{banks} provides a realization of a conformal sector
emerging from a renormalization flow toward the infrared below an
energy scale $\Lambda$ through dimensional transmutation, and
supersymmetric QCD may play a similar role  \cite{fox}.  Below this
scale there emerges an effective interpolating field, the unparticle
field,
 that features an anomalous scaling dimension  \cite{georgi}.

Various studies   recognized important
phenomenological  \cite{georgi,cheung,spin12,ira,Liao:2007ic}
astrophysical  \cite{raffelt,deshpande,freitas} and
cosmological  \cite{macdonald,davo,lewis,kame,he,kiku,xuelei,rich}
consequences of unparticles, including Hawking radiation into unparticles \cite{dai},
aspects of CP-violation \cite{zwicky1}, flavor physics \cite{flavor}
and low energy parity violation \cite{parity}. More recently, the consequences of mixing of
unparticle scalar fields and a Higgs field were studied in Ref. \cite{unparticleus} with
implications for slow-roll inflation.

A deconstruction program describes the unparticle as a tower of a continuum of excitations \cite{step}. However, this is not the only interpretation of unparticles. As mentioned
in Ref. \cite{unparticleus}, anomalous scaling dimensions are ubiquitous in critical phenomena near an infrared fixed point \cite{amit,brezin}, which is the main observation in Ref. \cite{banks}. There is also a well known phenomenon in QCD \cite{Neubert:2007kh} where anomalous dimensions emerge from
the multiple emission and absorption of gluons as a result from the resummation of infrared Sudakov
logarithms. Similarly, in QED, Bloch-Nordsieck resummation of infrared divergences arising from multiple emissions
and absorptions of photons yield threshold infrared divergences and  lead to a renormalized electron propagator that also features anomalous dimensions \cite{bloch,bogo}. The renormalization group resummation of absorption and emission of massless
quanta lead quite generally to anomalous dimensions in the propagators \cite{infradiv}.

A recent study \cite{liliu} suggests a connection between unparticles and the Miniboone anomaly, namely,
that the heaviest mass eigenstates corresponding to the mixed $\nu_\mu-\nu_e$ decays into the lightest
eigenstate and a scalar unparticle (see also \cite{Zhou:2007zq}). Furthermore, unparticle contributions to the neutrino-nucleon
cross section and its influence on the neutrino flux expected in a neutrino telescope such as IceCube
have been reported in Ref. \cite{telescopes}.

If heavy sterile neutrinos decaying into lighter active neutrinos and massless particles of a hidden conformal sector described by scalar unparticles are a plausible explanation
of the MiniBooNE anomaly, then the coupling of the sterile neutrinos to this degree of freedom will
necessarily lead to the consideration of the sterile neutrino \emph{itself} being an unparticle.

This consideration emerges naturally from the  ``deconstruction'' argument \cite{step}, since the
coupling of the sterile neutrino to the scalar unparticle leads to a spectral representation of the
sterile neutrino propagator
that features anomalous scaling dimensions. Alternatively, (but equivalently) the emission and absorption
of massless (conformal) quanta lead to infrared threshold divergences akin to Sudakov logarithms whose
renormalization group (or Bloch-Nordsieck \cite{bloch}) resummation leads to anomalous dimensions \cite{Neubert:2007kh,infradiv}.

Massive fermionic unparticles with a soft conformal breaking mass term had been introduced in
Ref. \cite{terning} and a very interesting proposal in which right handed neutrinos are fermionic
unparticles was considered in Ref. \cite{conformalquiros}.

If unparticle physics proves to be an
experimentally relevant extension of the standard model, it is
natural to consider sterile neutrinos, namely $SU(2)$ singlets,  as
being unparticles, and we refer to them as \emph{unsterile neutrinos}.

Unsterile neutrinos are assumed to couple to a   ``hidden'' conformal sector beyond the standard
model and acquire a (possibly large) anomalous scaling dimension below a scale $\Lambda$ of
dimensional transmutation at which the  infrared fixed point of the conformal sector dominates the
low energy dynamics.

In this article we consider such a possibility and study the
consequences of unsterile neutrinos mixing with active
neutrinos via a typical see-saw mass matrix. We consider the
simplest scenario of one unsterile and one active Dirac
neutrino to establish the general consequences of their mixing. Our
objectives in this article are two-fold:

\begin{itemize}
\item{ Because unsterile neutrinos feature non-canonical kinetic
terms, novel aspects of mixing phenomena emerge. We explore the
fundamental aspects of mixing   between these unsterile and
the usual active neutrinos via a see-saw type mass matrix.}

\item{We also focus on potential cosmological consequences, in particular  if and how
the unparticle nature of a sterile neutrino modifies its radiative
decay into an active-like neutrino. This decay rate is an important
ingredient to establish bounds on masses and mixing angles from the
cosmological X-ray or soft-gamma background in the case when the
mass of the sterile-like neutrino is in the keV range, which is of interest when considering it as a
dark matter candidate.}

\end{itemize}

Our results can be summarized as the following:

\begin{itemize}
\item{The spinor nature of the unparticle field introduces novel aspects of mixing, in which there is
no \emph{unitary} transformation purely in flavor space that diagonalizes the full propagator. The diagonalization
requires a non-unitary transformation and field redefinition followed by a momentum-dependent transformation that
is unitary below the unparticle threshold but non-unitary above. The
resulting mixing angles depend on the four-momentum\footnote{Ref. \cite{Schwetz:2007cd} speculated that fermionic unparticle coupled to an active neutrino might give rise to energy-dependent mixing.}. }

\item{ For a see-saw type mass matrix that mixes the unsterile and active neutrino, we find an unsterile-like mode with an ``invisible'' decay width. A renormalization group inspired ``resummation'' argument suggests that this width is
a result of the decay of the unsterile-like mode into the active-like mode and particles in the ``hidden
sector''. A complex pole for the unsterile mode exists only for $0 < \eta \leq 1/3$, where $\eta$ is the
unparticle anomalous dimension. As $\eta \rightarrow 1/3^-$ the real part of the pole approaches the
unparticle threshold from above and the width becomes large. For $\eta > 1/3$ the spectral density for
the unsterile mode does not feature a complex pole, but is described by a broad continuum with a large
enhancement at the unparticle threshold.

The active-like mode features a stable isolated pole below the unparticle threshold, but ``inherits'' a non-trivial spectral density above it as a consequence of the mixing, even in absence of standard model interactions. The non-vanishing spectral density may open up new kinematic channels for weak interactions.   }

\item{We obtain the \emph{radiative} decay width of the unsterile-like mode into the active-like mode and a photon via a charged current loop. For large anomalous dimension (but $\eta < 1/3$) we find a substantial suppression of the radiative decay width suggesting a concomitant weakening of the bounds on the mass and mixing angle  from the X-ray or soft gamma ray background.  }

\end{itemize}

\section{unsterile-active mixing}\label{sec:unmix}

In order to study the fundamental aspects of unsterile-active
neutrino mixing and the potential cosmological consequences, we consider the simplest case with one unsterile and one active Dirac
neutrino. The case of Majorana neutrinos and a triplet of unsterile
neutrinos as envisaged in extensions beyond the standard
model \cite{micha} will be studied in detail elsewhere.

For an unsterile Dirac fermion the Lagrangian density  in momentum
space is \cite{fox,terning}
\be \mathcal{L} = \overline{\psi}_U(-p)
\, \big(\pslash -M\big)F(p)\,\psi_U(p), \label{unlag}\ee
where
\be
F(p) = \Big[\frac{-p^2+M^2-i\epsilon}{\Lambda^2}\Big]^{-\eta}~~;~~ 0
\leq \eta < 1 \,.\label{F}\ee

$\Lambda$ is the scale below which the low energy dynamics is
dominated by the infrared fixed point of the conformal sector. Below
this scale the unparticle   is described by an interpolating
field whose two point correlation function scales with an anomalous
dimension \cite{georgi}. Consistency of the unparticle
interpretation requires that \be M < \Lambda \,.\label{MLambda}\ee

We consider the mixing with an ``active'' massless Dirac neutrino
$\psi_a$ of the form \be \mathcal{L}_m = \overline{\psi}_U \, m \,
\psi_a + \mathrm{h.c.} \label{mixlag}\ee  A see-saw mechanism
consistent  with the unparticle nature of the sterile neutrino,
namely an interpolating effective field below a scale $\Lambda$
entails the following hierarchy of scales \be m \ll M < \Lambda \,.
\label{hierarchy}\ee
In what follows we will explicitly invoke this
hierarchy in the analysis.

It is convenient to introduce the ``flavor doublet" \be \Psi =  \Big(\begin{array}{c}
                     \psi_a \\
                     \psi_U
                   \end{array}\Big), \label{doublet} \ee and write the Lagrangian density for the
unsterile and active fermions as
\be \mathcal{L} = \overline{\Psi}(-p)\,\Big[\pslash \, \mathds{F} - \mathds{M} \Big]\,\Psi(p), \label{mixed}\ee  where \be \mathds{F} = \Bigg( \begin{array}{cc}
                                                  1 & 0 \\
                                                  0 & F(p)
                                                \end{array}
\Bigg), \label{Fmat} \ee and
\be \mathds{M} =  \Bigg( \begin{array}{cc}
                                                  0 & m \\
                                                  m & M F(p)
                                                \end{array}
\Bigg) \label{Mmat}. \ee

The equation of motion is
\be \Big[\pslash \, \mathds{F} - \mathds{M} \Big]\,\Psi(p) =0. \label{eofm}\ee It is convenient to
introduce the spinor $\Phi$ so that
\be \Psi(p) = \Big[\pslash \, \mathds{I} + \mathds{F}^{-1}\, \mathds{M} \Big]\,\Phi(p), \label{nuspinor}\ee which obeys the following equation of motion
\be \Big[p^2 \mathds{F} - \mathds{M}~\mathds{F}^{-1}~\mathds{M} \Big]\Phi(p) =0, \label{eqphi}\ee where
\be   \mathds{M}~\mathds{F}^{-1}~\mathds{M} = \Bigg( \begin{array}{cc}
                                                                    \frac{m^2}{F(p)} & ~~ mM \\
                                                                    mM & ~~ m^2+M^2F(p)
                                                                  \end{array}
  \Bigg). \label{bigM}\ee

The matrix in (\ref{eqphi}) can be written as follows
  \be \Big[p^2 \mathds{F} - \mathds{M}~\mathds{F}^{-1}~\mathds{M} \Big] = \alpha(p)~\mathds{I}-\beta(p) \Bigg( \begin{array}{cc}
                                                                              -\mathcal{C}(p) & \mathcal{S}(p) \\
                                                                              \mathcal{S}(p) & \mathcal{C}(p)
                                                                            \end{array}
   \Bigg)\,. \label{mat2} \ee   Introducing the shorthand \be Q= p^2-\frac{m^2}{F(p)}\, , \label{Q}\ee
   we obtain

   \bea \alpha(p)  & = &  \frac{1}{2}\Bigg[ Q+ F(p)(Q-M^2) \Bigg]\,, \label{alpha}\\
   \beta(p) & = & \frac{1}{2} \Bigg[ \Big[Q-F(p)(Q-M^2) \Big]^2 +4 m^2 M^2\Bigg]^{\frac{1}{2}}\, ,\label{beta} \eea and
\bea \mathcal{C}(p) & = &  \frac{\Big[Q-F(p)(Q-M^2) \Big]}{2~\beta(p)}\, , \label{Cp}\\
\mathcal{S}(p) & = & \frac{  mM}{\beta(p)} \,.\label{Sp}\eea These
functions obey \be \mathcal{C}^2(p) + \mathcal{S}^2(p)  = 1 \,.
\label{sum2}\ee

It becomes clear that the dispersions relations for the propagating modes correspond to
\be \mathrm{det}\Big[p^2 \mathds{F} - \mathds{M}~\mathds{F}^{-1}~\mathds{M} \Big] =  \alpha^2(p)-\beta^2(p)  =0  \Rightarrow Q (Q-M^2)-\frac{m^2M^2}{F(p)} = 0, \label{det}\ee
which then leads to a self-consistent equation for the
dispersion relation of the propagating modes
\be p^2_{\pm} = \frac{m^2}{F(p_\pm)} + \frac{M^2}{2} \Bigg[1 \pm \sqrt{1+ \frac{4m^2}{M^2F(p_\pm)}}~\Bigg]. \label{ppm}\ee

The Klein-Gordon operator (\ref{mat2}), which we obtained from squaring the Dirac operator, can be diagonalized using the transformation
\be \mathcal{U}(p) = \Bigg( \begin{array}{cc}
                              c(p) & s(p) \\
                              -s(p) & c(p)
                            \end{array}
 \Bigg), \label{Uofp}\ee where
 \be c(p) = \frac{1}{\sqrt{2}} \Big[1+\mathcal{C}(p)\Big]^\frac{1}{2} ~~;~~ s(p) = \frac{1}{\sqrt{2}} \Big[1-\mathcal{C}(p)\Big]^\frac{1}{2}\, . \label{cands}\ee
The resulting spinor is given by
 \be \tilde{\Phi}(p)=\mathcal{U}^{-1}(p)\,\Phi(p) \,,\label{diagspin}\ee
and in this basis, the Klein-Gordon matrix (\ref{mat2}) becomes
\be \mathcal{U}^{-1}(p) \Big[p^2 \mathds{F} - \mathds{M}~\mathds{F}^{-1}~\mathds{M} \Big] \mathcal{U}(p) = \Bigg(\begin{array}{cc}
                         \alpha(p)+\beta(p) & 0 \\
                         0 & \alpha(p)-\beta(p)
                       \end{array}\Bigg). \label{mat2d} \ee

 An alternative manner to understand the solutions to the equations of motion is by going to the chiral
 representation. We do this by expressing the spinor $\Psi$ in terms of its right and left components, each a
 flavor doublet
 \be \Psi = \Bigg(\begin{array}{c}
              \Psi_R \\
              \Psi_L
            \end{array}\Bigg). \ee
We can then expand  the right and left components in the helicity basis
\bea \Psi_{R,L} & = &
\sum_{h=\pm 1}  v^h(\vp) \otimes \xi^h_{R,L}\, ,  \label{PsiRL} \eea
where
\be\frac{ \vec{\sigma} \cdot \vp }{|\vp|}~   v^h(\vp) = h ~ v^h(\vp)
~~;~~ h= \pm 1, \label{helexpansion}\ee
and $\xi^h_{R,L}$ are
 flavor doublets. We find that
\bea (p^0 - h |\vp|) \mathds{F}~\xi^h_R + \mathds{M}~ \xi^h_L & = & 0, \label{Req}\\
(p^0 + h |\vp|) \mathds{F}~\xi^h_L + \mathds{M}~ \xi^h_R & = & 0
\,.\label{Leq}\eea
Using (\ref{Leq}), we can expres $\xi^h_L$ in terms of $\xi^h_R$ and obtain
\be \Bigg[p^2 \mathds{F}-
\mathds{M}~\mathds{F}^{-1}~\mathds{M} \Bigg]\xi^h_R =0 ~~;~~ \xi^h_L
= -\frac{\mathds{F}^{-1}~\mathds{M}}{(p^0+h|\vp|)} ~\xi^h_R\, .
\label{Req2L}\ee
Alternatively, we can express $\xi^h_R$ in terms of
$\xi^h_L$ using (\ref{Req}), and obtain \be \Bigg[p^2 \mathds{F}-
\mathds{M}~\mathds{F}^{-1}~\mathds{M} \Bigg]\xi^h_L =0 ~~;~~ \xi^h_R
= -\frac{\mathds{F}^{-1}~\mathds{M}}{(p^0-h|\vp|)} ~\xi^h_L \, .
\label{Leq2R}\ee

Introducing the ``mass eigenstates'' \be \chi^h_{R,L} = \mathcal{U}^{-1}(p)~ \xi^h_{R,L}\, , \label{masseigen}\ee it follows
that \be \Bigg(\begin{array}{cc}
                         \alpha(p)+\beta(p) & 0 \\
                         0 & \alpha(p)-\beta(p)
                       \end{array}\Bigg) \chi^h_R = 0 ~~;~~ \chi^h_L = -\frac{\mathcal{U}^{-1}(p)\Big(\mathds{F}^{-1}~\mathds{M}\Big)\mathcal{U}(p)}{(p^0+h|\vp|)}
                        ~\chi^h_R\, , \label{masseigenR} \ee or
\be \Bigg(\begin{array}{cc}
                         \alpha(p)+\beta(p) & 0 \\
                         0 & \alpha(p)-\beta(p)
                       \end{array}\Bigg) \chi^h_L = 0 ~~;~~ \chi^h_R = -\frac{\mathcal{U}^{-1}(p)\Big(\mathds{F}^{-1}~\mathds{M}\Big)\mathcal{U}(p)}{(p^0-h|\vp|)}
                        ~\chi^h_L\, . \label{masseigenL} \ee

It is clear that although the transformation $\mathcal{U} $
diagonalizes the Klein-Gordon operator, for $F \neq 1$, it does \emph{not} diagonalize the
flavor matrix $\mathds{F}^{-1}\mathds{M}$ in (\ref{masseigenR}) and
(\ref{masseigenL}).

The dispersion relations of the propagating eigenstates correspond to the solutions of $\alpha(p) = \pm \beta(p)$, which
are determined by the self-consistent equation (\ref{ppm}). The roots $p_{\pm}$ correspond to $\alpha(p_{\pm}) = \pm \beta(p_{\pm})$ respectively.

The propagator for the flavor doublet $\Psi$, denoted by $\mathds{S}$, obeys \be  \Big[\pslash \, \mathds{F} -
\mathds{M} \Big]\,\mathds{S}(p) = \mathds{I}, \label{prop}\ee where
$\mathds{I}$ is the identity in both flavor and Dirac space.
Pre-multiplying (\ref{prop}) by $\Big[\pslash \, \mathds{I} + \mathds{M}\,
\mathds{F}^{-1}    \Big]$, we obtain \be  \Big[p^2
\mathds{F} - \mathds{M}~\mathds{F}^{-1}~\mathds{M} \Big]\,
\mathds{S}(p) = \Big[\pslash \, \mathds{I} + \mathds{M}\, \mathds{F}^{-1}
  \Big]. \label{prop2}\ee
In the new basis $\tilde{\Psi}(p)=\mathcal{U}^{-1}(p)\,\Psi(p)$, the propagator is expressed by
\be
\mathcal{U}^{-1}(p)\,\mathds{S}(p)\, \mathcal{U}(p) = \Bigg(
                                                        \begin{array}{cc}
                                                          \frac{1}{\alpha(p)+\beta(p)} & 0 \\
                                                          0 & \frac{1}{\alpha(p)-\beta(p)} \\
                                                        \end{array}
                                                      \Bigg)\,\Big[\pslash \, \mathds{I} + \mathcal{U}^{-1}(p)\Big(\mathds{M}\, \mathds{F}^{-1} \Big)
  \mathcal{U}(p) \Big] \,.\label{propdiag}\ee Only for
$F=1$ is the matrix inside the bracket on the right hand side of
(\ref{propdiag}) diagonal. For $F\neq 1$, there is \emph{no
unitary transformation} that diagonalizes both the matrices
proportional to $\pslash$ and  $  \mathds{M}\, \mathds{F}^{-1}  $.

\emph{If} $F(p)$ is real, $\mathcal{C}(p),\mathcal{S}(p)$ introduced
in (\ref{mat2},\ref{Cp},\ref{Sp}) can be identified as cosine and sine of (twice) of the \emph{mixing
angle}, namely \be \mathcal{C}(p) = \cos(2\theta_m(p))
~~;~~\mathcal{S}(p) = \sin(2\theta_m(p)), \label{mixangs} \ee
and similarly $c(p),s(p)$ in (\ref{Uofp},\ref{cands}) \be c(p) = \cos(\theta_m(p))~~;~~ s(p) = \sin(\theta_m(p)). \label{cosin}\ee

Therefore, \emph{if} $F(p)$ is real, the transformation
$\mathcal{U}(p)$ is unitary and $\theta_m(p)$ is identified
as the mixing angle. However, $F(p)$ becomes complex above threshold
$p^2 > M^2$ reflecting the multiparticle nature of the unparticle
interpolating field $\psi_U$. Thus, $\mathcal{C},\mathcal{S},c$ and $s$
cannot be interpreted as cosine and sine of (twice) the mixing
angle.

For $F(p)=1$, the unparticle field is just an ordinary Dirac
spinor field with canonical kinetic term. In this case, $ \mathcal{C},\mathcal{S},c$ and $s$ become independent
of $p$, and they are given by \be \mathcal{C} = \cos(2\theta_0)= \frac{M}{\sqrt{M^2+4m^2}}~~;~~\mathcal{S} = \sin(2\theta_0)=\frac{2m}{\sqrt{M^2+4m^2}}\, , \label{vacmixan}\ee and \be c=\cos(\theta_0)~~;~~s = \sin(\theta_0). \label{angsvac}\ee
Here, the angle
$\theta_0$ is the usual mixing angle for the see-saw
mass matrix.

Although the transformation  $\mathcal{U}$ (\ref{Uofp}) diagonalizes the Klein-Gordon operator
in the equations of motion (\ref{mat2d}), it
does \emph{not} diagonalize the propagator or the Lagrangian in terms of the ``mass eigenstates''.
Furthermore, the solutions of the equations of motion in the transformed basis, namely the spinor
$\chi$ (\ref{masseigenR},\ref{masseigenL}), still has mixing between them. This is because $\mathcal{U}^{-1}(p)\Big(\mathds{F}^{-1}~\mathds{M}\Big)\mathcal{U}(p)$ is not
diagonal. It follows that there is \emph{no unitary transformation that diagonalizes the propagator}.
The problem in the diagonalization can be traced to the spinor nature of the unparticle field, where there
are two independent structures in the effective action, the mass matrix term and the kinetic term  $\pslash$ multiplied by $\mathds{F}$. There simply is no unitary transformation that diagonalizes
simultaneously both the mass term $\mathds{M}$ and the kinetic term $\mathds{F}$. An alternative explanation using Lorentz invariance argument is presented in the Appendix \ref{Alt}.

\section{Non-unitary transformation: Field redefinition \label{mixing}}

As pointed out in the previous section because of the spinor nature of the field and the fact
that the matrix coefficients of the kinetic term $\pslash$ and the mass matrix do not commute, there
is no unitary transformation that diagonalizes the full propagator, even for real $F(p)$. However,
the equation for the propagator (\ref{prop}) suggests that the following set of transformations will lead
to a diagonalization of the propagator. Let us introduce \be \widetilde{\mathds{S}} = \sqrt{\mathds{F}} ~\mathds{S}~
\sqrt{\mathds{F}}\,. \label{Sredef}\ee
By multiplying the equation (\ref{prop}) on the right  by $\sqrt{\mathds{F}}$ and on the left by $1/\sqrt{\mathds{F}}$ one finds the following equation for $\widetilde{\mathds{S}}$
\be \Big[p^2 - \widetilde{\mathds{M}}^2\Big] \widetilde{\mathds{S}} = \pslash +\widetilde{\mathds{M}},  \label{direqred}\ee where   \be    \widetilde{\mathds{M}} = \frac{1}{\sqrt{\mathds{F}}} ~ \mathds{M} ~ \frac{1}{\sqrt{\mathds{F}}} = \Bigg( \begin{array}{cc}
         0 & \frac{m}{\sqrt{F(p)}} \\
         \frac{m}{\sqrt{F(p)}} & M
       \end{array}
 \Bigg). \label{redmass}\ee

 The mass matrix $\widetilde{\mathds{M}}$ can be written as
 \be \widetilde{\mathds{M}} = \frac{M}{2}~  \mathds{I} + \frac{M}{2} \Bigg[1+ \frac{4 m^2}{M^2 F(p)} \Bigg]^\frac{1}{2}~~\Bigg(\begin{array}{cc}
                             -\widetilde{\mathcal{C}}(p) & \widetilde{\mathcal{S}}(p) \\
                             \widetilde{\mathcal{S}}(p) & \widetilde{\mathcal{C}}(p)
                           \end{array}
   \Bigg), \label{massmatred}\ee
where
\bea \widetilde{\mathcal{C}}(p) & = &  \Bigg[1+ \frac{4 m^2}{M^2 F(p)} \Bigg]^{-\frac{1}{2}} \,,\label{tilC}\\
\widetilde{\mathcal{S}}(p) & = & \frac{2 m}{M ~ \sqrt{F(p)}}~
\Bigg[1+ \frac{4 m^2}{M^2 F(p)}\Bigg]^{-\frac{1}{2}}\,. \label{tilS}\eea
When $F(p)$ is real
 \be \widetilde{\mathcal{C}}(p) = \cos(2\varphi(p)) ~~;~~\widetilde{\mathcal{S}}(p)   \sin(2\varphi(p)), \label{mixangred}\ee where $\varphi(p)$ is a mixing angle that depends on $p^2$.

It is clear that now the propagator can be diagonalized by the matrix
\be   U(p) = \left(
                      \begin{array}{cc}
                        \widetilde{c}(p) & \widetilde{s}(p) \\
                        -\widetilde{s}(p) & \widetilde{c}(p) \\
                      \end{array}
                    \right), \label{Uofp}\ee where
 \be \widetilde{c}(p) = \frac{1}{\sqrt{2}} \Big[1+\widetilde{\mathcal{C}}(p)\Big]^\frac{1}{2} ~~;~~ \widetilde{s}(p) = \frac{1}{\sqrt{2}} \Big[1-\widetilde{\mathcal{C}}(p)\Big]^\frac{1}{2}\,. \label{candsred}\ee The mass matrix is now diagonal
 \be U^{-1}(p) ~\widetilde{\mathds{M}}~U(p) = \widetilde{\mathds{M}}_d= \Bigg(\begin{array}{cc}
                                                      M_1(p) & 0 \\
                                                      0& M_2(p)
                                                    \end{array}
  \Bigg), \label{Mdiagred}\ee with \bea M_1(p) & = & \frac{M}{2} \Bigg[1- \Bigg(1+ \frac{4 m^2}{M^2 F(p)} \Bigg)^{\frac{1}{2}} \Bigg], \label{M1red}\\ M_2(p) & = & \frac{M}{2} \Bigg[1+ \Bigg(1+ \frac{4 m^2}{M^2 F(p)} \Bigg)^{\frac{1}{2}} \Bigg]. \label{M2red} \eea

 If $F(p)$ is real, it follows that \be \widetilde{c}(p) = \cos(\varphi(p))~~;~~\widetilde{s}(p) = \sin(\varphi(p))\,. \label{cossinred}\ee The transformed propagator \be \widetilde{\mathds{S}}_m  U^{-1}(p)~\widetilde{\mathds{S}}~U(p) \label{tranS}\ee is given by
 \be \widetilde{\mathds{S}}_m  = \Bigg( \begin{array}{cc}
                                          \frac{\pslash ~ + M_1(p)}{p^2-M^2_1(p)} & 0 \\
                                          0 & \frac{\pslash ~ + M_2(p)}{p^2-M^2_2(p)}
                                        \end{array}
   \Bigg)\,. \label{dirdiag} \ee

The transformation (\ref{Sredef}) has a natural interpretation in terms of a \emph{field redefinition}. This can be inferred from the form of the kinetic term for the unparticle field $\overline{\psi}_U \pslash ~F(p) ~\psi_U$,  which suggests
that $F(p)$ can be interpreted as a momentum dependent  wave function renormalization. Let us define the rescaled
field as
\be \nu_U = \sqrt{F(p)}~~ \psi_U, \label{unfieldred}\ee
which along with  \be \nu_a \equiv \psi_a, \label{acred}\ee forms the flavor doublet
\be \nu = \Bigg(\begin{array}{c}
             \nu_a \\
             \nu_U
           \end{array} \Bigg) \,.\label{doubletred}\ee
With this field redefinition, the Lagrangian density becomes \be \mathcal{L} = \overline{\nu}
\big[\pslash \,\mathds{I} - \widetilde{\mathds{M}}\big] \nu \,.
\label{redefL}\ee
 No physics has been lost with this field redefinition as the correlation functions
 of the original unparticle field $\psi_U$ may be obtained as follows. Let us introduce Grassman sources $\mathcal{J}_U$
 coupled to the unparticle field in the Lagrangian density, namely
 \be \mathcal{L} \rightarrow \mathcal{L}+ \overline{\psi}_U \mathcal{J} + \mathrm{h.c}. \ee
 Upon field redefinition
 (\ref{unfieldred}), the source terms become $\overline{\nu}_U \mathcal{J}/\sqrt{F(p)}$, etc. Furthermore,
 at the level of the path integral, the field redefinition multiplies the measure by an overall field
 independent constant which cancels in all correlation functions.

 The necessity of a field redefinition to rescale to unity the
 coefficient of $\pslash$ has been also recognized in
 Ref. \cite{machet} within the context of radiative corrections in
 the quark sector of the standard model.

The full Lagrangian density can now be diagonalized by introducing the ``mass basis'' $\nu_1,\nu_2$
as
\be \Bigg( \begin{array}{c}
            \nu_a\\
             \nu_U
           \end{array}
\Bigg) =   U(p)   \Bigg( \begin{array}{c}
             \nu_1 \\
             \nu_2
           \end{array}
\Bigg) \,,  \label{unitrafred}\ee where $U(p)$ is given by Eq. (\ref{Uofp}).

 The dispersion relations are obtained from the respective Dirac equations for the mass eigenstates,
 \be \big[\pslash - M_i \big] \nu_i = 0 ~~;~~i=1,2 \label{Diracred}\ee leading to the self-consistent equations \bea p^2_1 & = & M^2_1(p_1) \frac{m^2}{F(p_1)} + \frac{M^2}{2}  \Bigg[1- \Bigg(1+ \frac{4 m^2}{M^2 F(p_1)} \Bigg)^{\frac{1}{2}}~ \Bigg], \label{p1}\\
 p^2_2 & = & M^2_2(p_2)  \frac{m^2}{F(p_2)} + \frac{M^2}{2}  \Bigg[1+ \Bigg(1+ \frac{4 m^2}{M^2 F(p_2)} \Bigg)^{\frac{1}{2}}~ \Bigg]. \label{p2} \eea
 These dispersion relations are exactly the same as Eq. (\ref{ppm}) with $p_{1,2}=p_{-,+}$ respectively. The reason for
 this is that the determinant (\ref{det}), which determines the dispersion relation, is simply rescaled, namely \be  \mathrm{det}\Big[p^2 \mathds{F} - \mathds{M}~\mathds{F}^{-1}~\mathds{M} \Big]   = \mathrm{det}\Bigg[ \sqrt{\mathds{F}}\Big(p^2 ~\mathds{I}  - \widetilde{\mathds{M}}^2  \Big) \sqrt{\mathds{F}}\Bigg]   = F(p) ~\mathrm{det}\Big[   p^2 ~\mathds{I}  - \widetilde{\mathds{M}}^2    \Big]\,. \label{detred}\ee

  The original unsterile and active fields are related to the mass eigenstates $\nu_{1,2}$ as
  \bea   \psi_a & = &  \Big[\widetilde{c}(p)\nu_1 +\widetilde{s}(p) \nu_2\Big], \label{psia}\\
  \psi_U & = & \frac{1}{\sqrt{F(p)}}\Big[\widetilde{c}(p)\nu_2 -\widetilde{s}(p) \nu_1\Big]\,. \label{psiU}\eea
Therefore, arbitrary correlation functions of the unsterile field $\psi_U$ can be
  obtained from the propagators of the mass eigenstates $\nu_1,\nu_2$. Furthermore, since the unsterile
  field does not couple to any other field of the standard model, only the unsterile-like mass eigenstate
  $\nu_2$ can participate in weak interaction processes via the relation (\ref{psia}) and this field
  does not directly involve the field redefinition (\ref{unfieldred}).

  For $p^2 < M^2$,  $F(p)$ is real and the transformation $U(p)$ is unitary.
  $U(p)$ is determined by the mixing angles defined by
  (\ref{tilC},\ref{tilS},\ref{mixangred}). However, above the
  unparticle threshold $p^2 > M^2$,  $F(p)$ is complex and the
  matrix $U(p)$ is \emph{not} unitary. This is a consequence of the coupling to a continuum of states.
  A similar situation emerges in the theory of neutral meson mixing,
  where the absorptive part of the Wigner-Weisskopf Hamiltonian, which
  describes the quantum mechanics of neutral meson mixing, prevents a
  diagonalization of the Hamiltonian via a unitary
  transformation. This situation has been
  analyzed in detail in Refs. \cite{beuthe,silva}. In particular,
  Ref. \cite{silva} discusses the reciprocal basis that
  corresponds to fields that are transformed by a non-unitary
  transformation. A similar discussion appropriate to the quark
  sector of the standard model is given in Ref. \cite{machet} and a non-unitary transformation concerning time-reversal violation in
  the neutral kaon system can be found in Ref. \cite{alvarez}. The reader is referred to these references for a detailed
  discussion of the reciprocal (or dual or in-out) basis within the theory of neutral meson mixing.

  The same analysis in terms of the reciprocal (or dual) basis applies to the case under consideration for
  $p^2 > M^2$ when $F(p)$ becomes complex.

\subsection{Complex Poles and Spectral Densities for the Active-like Mode}
 The dispersion relations of the propagating modes, are obtained from the complex
 poles of the propagator corresponding to $p^2_{1,2} = M^2_{1,2}$.   Self-consistent solutions of the equations (\ref{p1},\ref{p2}) are
 in general difficult to obtain analytically, however progress can be made in the relevant case
 $m\ll M$ and assuming self-consistently that $m^2/M^2 F(p_{1,2}) \ll 1$.

 With this approximation we find \be M^2_1(p) =  M^2\Bigg[\frac{m^4}{M^4\,F^2(p )}+\cdots \Bigg] \,, \label{M21} \ee therefore the self-consistent equation (\ref{p1}) for $p^2_1$ becomes
\be p^2_1  =   M^2\Bigg[\frac{m^4}{M^4\,F^2(p_1)}+\cdots \Bigg]\,. \label{pminus} \ee  Anticipating
self-consistently that $p^2_1 \ll M^2$ we write \be F(p_1) = \Bigg[\frac{M^2}{\Lambda^2} \Bigg]^{-\eta}
\,\Bigg[1-\frac{p^2_1}{M^2}\Bigg]^{-\eta}\,, \ee leading, to lowest order in the ratio $m^2/M^2F$, to the
solution
\be p^2_1 = \frac{m^4}{M^2} \Bigg[\frac{M^2}{\Lambda^2} \Bigg]^{2\eta} \equiv M^2_1, \label{p2min}\ee namely an isolated pole   below  the multiparticle threshold at $p^2 = M^2$ .
Near this pole  we find \be \frac{1}{p^2 - M^2_1(p)} \approx \frac{Z_1}{p^2 - M^2_1}, \label{alfaplusbeta}\ee where  \be Z^{-1}_1 \approx   1+ 2\eta ~ \frac{M^2_1}{M^2}  \,. \label{Z1}\ee

 For $F(p)=1$ ($\eta =0$), $M_1 = m^2/M $ is recognized as the smallest eigenvalue of the see-saw mass
  matrix, namely the mass of the lightest neutrino. This pole lies on  the real $p^2$ axis and describes a stable active-like propagating mode.

\begin{figure}[h]
\begin{center}
\includegraphics[width=10cm,keepaspectratio=true]{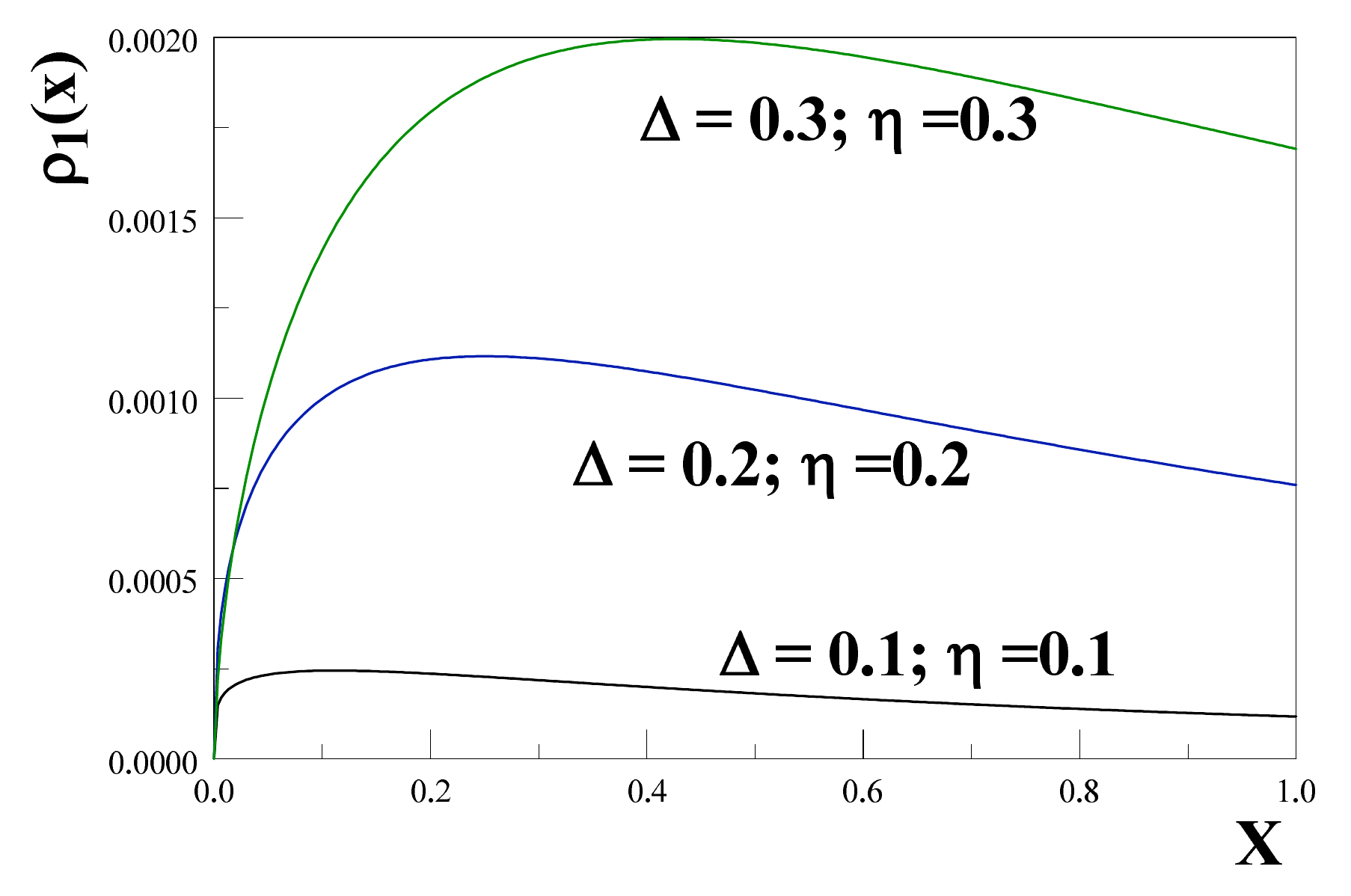}
\caption{Spectral density for the active-like mode.}
\label{fig:rho1}
\end{center}
\end{figure}

  The active-like propagator also features a discontinuity  across the real axis in the complex $p^2$-plane for $p^2 > M^2$, since \be p^2-M^2_1(p) =  p^2 - \frac{m^4}{M^2} ~ \Big[\frac{-p^2+M^2-i\epsilon}{\Lambda^2}\Big]^{2\,\eta} \,.\label{invprop1} \ee
  It is convenient to introduce the dimensionless variables
  \be x = \frac{p^2-M^2}{M^2} ~~;~~\Delta = 2~\frac{m^2}{M^2} \Bigg[\frac{M^2}{\Lambda^2} \Bigg]^{\eta}\,, \label{dimvars}\ee
   and use these to define the dimensionless spectral density
   \be \rho_1(x) = \frac{M^2}{2\pi i} ~ \mathrm{Disc}~\Big(\frac{1}{p^2 - M^2_1(p)}  \Big) \,,\label{rho1} \ee
   where the discontinuity is non-vanishing for $x >0$. We find that
    \be \rho_1(x) = \frac{\Theta(x)}{\pi} ~\frac{\frac{\Delta^2}{4}\,x^{2\eta}\sin(2\pi\eta)}
    {\Big[ x+1 - \frac{\Delta^2}{4}\,x^{2\eta}\cos(2\pi\eta) \Big]^2+\Big[\frac{\Delta^2}{4}\,x^{2\eta}\sin(2\pi\eta) \Big]^2}\, .  \label{disc1}\ee
     This spectral density vanishes at threshold $p^2=M^2$ ($x=0$), increases rapidly reaching a broad maximum and diminishes
      for increasing $x$ (Fig. (\ref{fig:rho1})).

It is remarkable that in the absence of other interactions, the
propagator of the active-like (lightest) mass eigenstate features a
non-vanishing spectral density away from its mass shell for $p^2 >
M^2 > M^2_1$. This mode ``inherits'' a coupling to the continuum
``hidden'' sector as a consequence of the mixing with the
unparticle. The non-vanishing spectral density above the unparticle threshold at $p^2 = M^2$ may lead
to opening new kinematic channels when the active-like neutrino is coupled to the standard model fields.

\subsection{Complex Poles and Spectral Densities for the Unsterile-like Mode}

 For $m\ll M$, the self-consistent equation  (\ref{p2}) for  the unsterile-like mode      becomes
 \be  p^2_2   -    M^2 = {2m^2}\Big[\frac{-p^2_2+M^2-i\epsilon}{\Lambda^2}\Big]^{ \eta} +\cdots \label{pp2}\ee
In terms of the dimensionless variables (\ref{dimvars}) this equation becomes \be x = \Delta \Big[-x -i\epsilon\Big]^\eta \,.\label{xeqn}\ee We find that there is a solution \emph{only} for $$\mathrm{Re}(x) > 0~,~ 0 \leq \eta < 1/3,$$ and it is given by
  \be p^2_2 = M^2 \Big[1 + \Delta^\frac{1}{1-\eta}\,\Big\{\cos\Big( \frac{\pi \eta}{1-\eta}\Big)-i \sin\Big( \frac{\pi \eta}{1-\eta}\Big) \Big\}\Big]\,. \label{solplus}\ee This solution describes a pole in the complex plane (a resonance) and near this pole we find
  \be \frac{1}{p^2-M^2_2(p)} \approx \frac{Z_2}{p^2-M^2_2+iM_2\Gamma}, \label{alfaminbeta}\ee where
  \be M^2_2 = M^2 \Big[1 + \Delta^\frac{1}{1-\eta}\, \cos\Big( \frac{\pi \eta}{1-\eta}\Big)\Big]\;,\label{M22}\ee
  \be \Gamma =\frac{ M^2}{M_2}   \Delta^\frac{1}{1-\eta}\,\sin\Big( \frac{\pi \eta}{1-\eta}\Big) \;,\label{gama} \ee and\footnote{We note that the unparticle field is \emph{not canonical}, therefore the
  residue at  the unparticle-like ``pole'' is \emph{not restricted } by canonical commutation relations to obey $0< Z_2 \leq 1$.}
  \be Z_2 = \frac{1}{1-\eta}\,. \label{Z2}\ee

  The imaginary part is a consequence of the fact that the real part of the pole is above
  the unparticle continuum determined by the multiparticle threshold at $p^2 = M^2$. For
  $F(p)=1$ ($\eta =0$), the largest eigenvalue of the see-saw mass matrix is at $M+ m^2/M > M$.
  After mixing, the new pole is in the unparticle continuum, moving off the real axis into
  a second (or higher) Riemann sheet in the complex $p^2$ plane. The imaginary part describes the
  \emph{decay} of the unsterile like mode into the active-like mode and particles in the ``hidden''
  conformal sector. A similar phenomenon was observed in the bosonic case in Ref. \cite{unparticleus}.  We refer
  to $\Gamma$ as the ``invisible width'' of the unsterile-like neutrino since it describes its decay into
  an active-like and conformal massless particles in the ``hidden sector.'' A more detailed discussion and interpretation of this result, based on a renormalization group \cite{infradiv,bogo,Neubert:2007kh} resummation is presented in Sec. \ref{resum}.

\begin{figure}[h]
\begin{center}
\includegraphics[width=10cm,keepaspectratio=true]{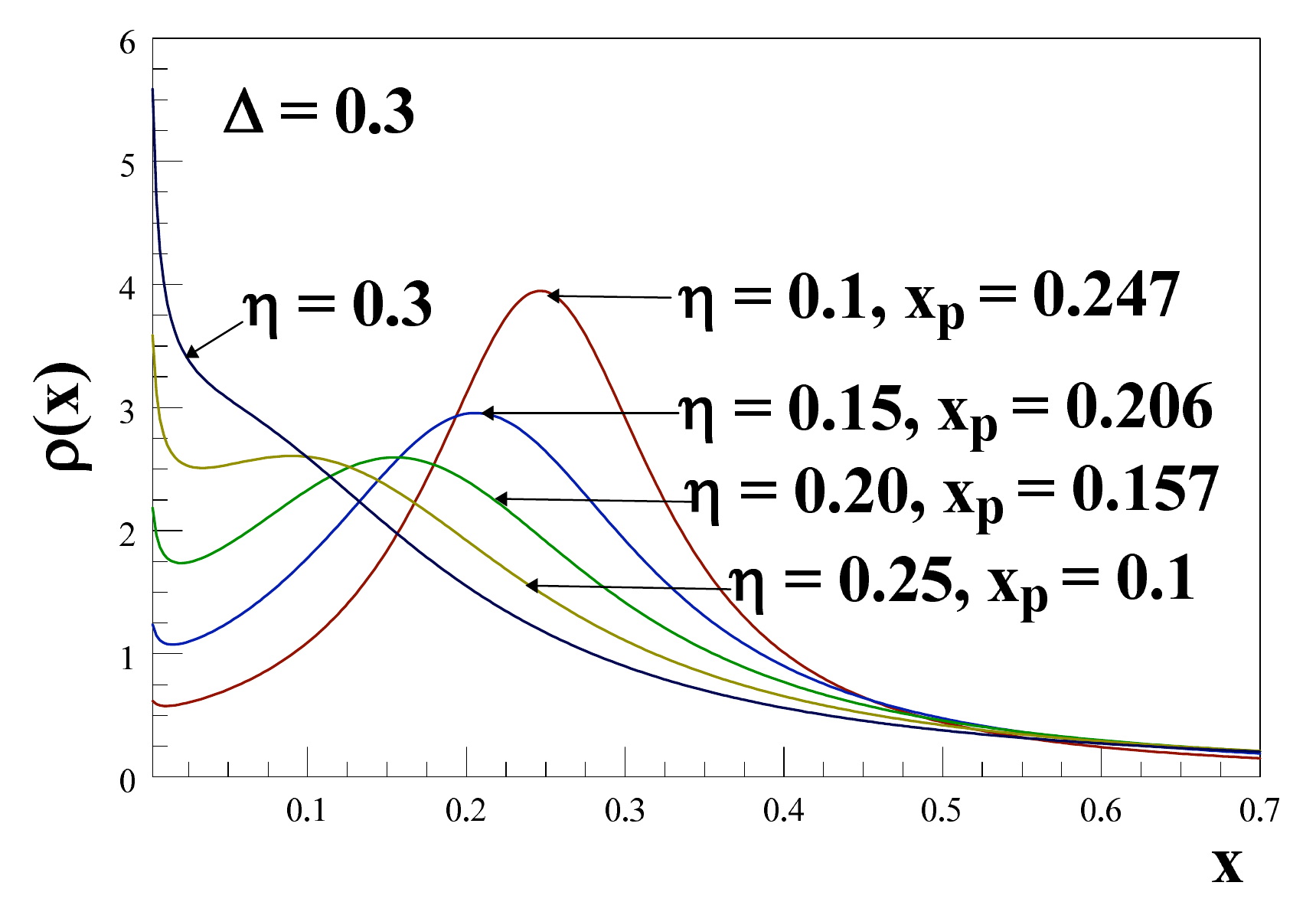}
\caption{$\rho_2(x)$ vs. $x$ for $\Delta = 0.3$ and several values of   $\eta$. $x_p$ is the real
part of the complex pole (\ref{M22},\ref{xpole}).}
\label{fig:rho2}
\end{center}
\end{figure}

   The   spectral density is obtained from the discontinuity across the real axis
  in the complex $p^2$ plane (see Eq.(\ref{rho1}))
  \be \rho_2(x) = \frac{\Theta(x)}{\pi} ~\frac{ \Delta \,x^{\eta}\sin(\pi\eta)}{\Big[ x  - {\Delta }\,x^{\eta}\cos(\pi\eta) \Big]^2+\Big[{\Delta}\,x^{\eta}\sin(\pi\eta) \Big]^2}\,,  \label{disc2}\ee which is displayed in Fig. (\ref{fig:rho2}). For $0 \leq \eta < 1/3$, there is a resonance with a
  maximum confirmed to be given by \be x_p = \Delta^\frac{1}{1-\eta}\, \cos\Big( \frac{\pi \eta}{1-\eta}\Big), \label{xpole}\ee as obtained in Eq. (\ref{M22}).

  For $\Delta \ll 1$ the real part of the unsterile-like pole is very near the threshold at $p^2 = M^2$ and therefore the dimensionless  ratio \be \frac{\Gamma}{M_2-M} \simeq  \frac{ 2 \tan \Big( \frac{\pi \eta}{1-\eta}\Big) }{\Big[1 + \Delta^\frac{1}{1-\eta}\, \cos\Big( \frac{\pi \eta}{1-\eta}\Big)\Big]^\frac{1}{2}} \label{ratio}\ee determines if the resonance is
  broad or narrow as compared to the distance between threshold and the resonance.

   For $\eta \ll  1$ ($F(p) \rightarrow 1$),  the width becomes very small
  and the resonance is sharp, centered at the mass of the sterile neutrino $M+m^2/M$. As $\eta \rightarrow 1/3$ from below, the real part of the pole
  approaches threshold ($x_p \rightarrow 0$) while the width remains constant. The resonance
  broadens enormously since the ratio (\ref{ratio}) diverges as $\eta \rightarrow 1/3^-$, and merges with the threshold at $\eta = 1/3$. There are no solutions of the
  self-consistency condition (\ref{pp2}) for a complex pole for $\eta > 1/3$.

  \subsection{Resummation Interpretation of the Decay Width \label{resum}}

\begin{figure}[h]
\begin{center}
\includegraphics[width=6 cm,keepaspectratio=true]{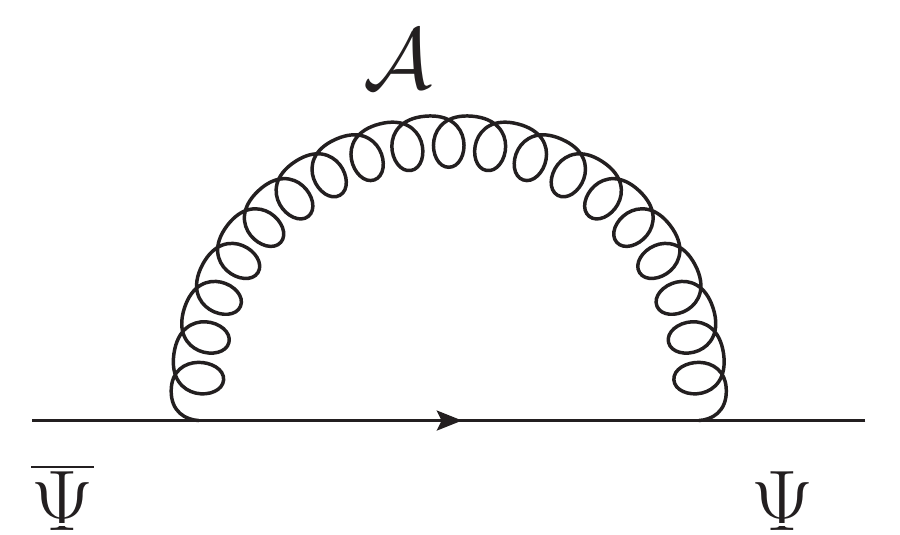}
\caption{$\Psi$ self-energy to lowest order in g. $\mathcal{A}$ is the conformal field.}
\label{fig:selfenergy}
\end{center}
\end{figure}

  Consider the unparticle
  to be a single Dirac fermion (neutrino) interacting with massless
  particles of the conformal hidden sector, described by a conformal field $\mathcal{A}$. Let us
  assume such interaction to be of the form \be \mathcal{L}_{int} = g\, \overline{\Psi} \mathcal{A} \Psi, \label{lint}\ee where $g$ is a small dimensionless coupling. In perturbation theory, the self-energy
  of the ``unparticle field'' $\Psi$ is depicted in Fig. (\ref{fig:selfenergy})

 To lowest order in g, the self energy, which is once-subtracted to vanish at $\pslash = M$, is given near the mass shell by \be \Sigma (p) = -\eta (\pslash -M)\,\ln\Bigg[\frac{-p^2+M^2-i\epsilon}{\Lambda^2}\Bigg], \label{selfenergy} \ee where $\Lambda$ is
 a renormalization scale,
 \be \eta = c g^2,\label{eta}\ee and $c$ is a constant that depends on the
 nature of the conformal field $\mathcal{A}$ (gauge or scalar massless particle).

  Integrating out the conformal field $\mathcal{A}$ leads to the
 following effective action for $\Psi$
\bea
\mathcal{L}_{eff} &=& \overline{\Psi}(\pslash -M)\Bigg[1-\eta
 \ln\Big(\frac{-p^2+M^2-i\epsilon}{\Lambda^2}\Big)\Bigg] \Psi \nonumber \\
 &\approx& \overline{\Psi}(\pslash -M)\Bigg[ \frac{-p^2+M^2-i\epsilon}{\Lambda^2}\Bigg]^{-\eta} \Psi, \label{Leff}
 \eea where in the last line we have invoked a renormalization group  resummation \cite{infradiv,bogo,Neubert:2007kh}  of the infrared threshold divergences.
The infrared logarithmic
 divergence at $p^2 = M^2$ and the imaginary part for $p^2 > M^2$ is the result of the emission and
 absorption of massless quanta, an ubiquitous phenomenon in gauge theories (for a discussion within QCD, see \cite{Neubert:2007kh}).

\begin{figure}[h]
\begin{center}
\includegraphics[width=6cm,keepaspectratio=true]{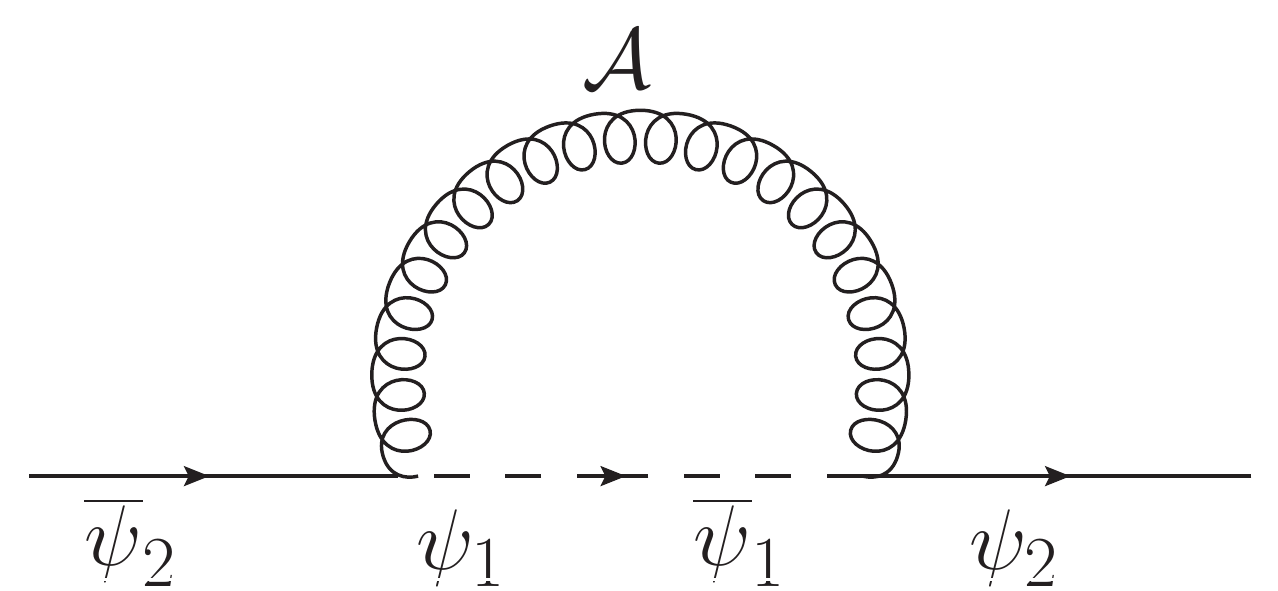}
\caption{Self-energy for mass eigenstate $\psi_2$.}
\label{fig:se12}
\end{center}
\end{figure}

 Now consider coupling the heavy neutrino $\Psi$ to a massless (active) neutrino $\psi_a$ via the
 coupling \be \mathcal{L}_m = \overline{\Psi} ~ m ~\psi_a + \mathrm{h.c.} \,,\label{coup2} \ee  leading
 to a see-saw mass matrix of the same form as in Eq. (\ref{Mmat}) for $F(p)=1$ with $m\ll M$. The see-saw mass matrix can be diagonalized by a unitary transformation with the mixing angle $\theta_0$, given by the  relations (\ref{vacmixan}). The fields that describe the ``mass eigenstates'' are
 \bea \psi_1 & = &  \cos(\theta_0) \, \psi_a + \sin(\theta_0) \, \Psi, \label{psi1}\\
 \psi_2 & = &  \cos(\theta_0)\, \Psi - \sin(\theta_0)\, \psi_a\,, \label{psi2} \eea where
 the masses corresponding to the fields $\psi_{1,2}$ are $M_{1,2}$ respectively, with
 \be M_1 \approx \frac{m^2}{M}~~;~~M_2 \approx M+ \frac{m^2}{M}\,. \label{masses12}\ee To lowest order in the
 see-saw ratio $m/M$, it follows from (\ref{vacmixan}) that
 \be \sin(\theta_0) = \frac{m}{M}\,. \label{sino}\ee The $\Psi-\psi_a$ mixing leads to the
 following interaction vertex between the fields associated with the mass eigenstates and the conformal
 field $\mathcal{A}$
 \be \mathcal{L}_{int} = g\,\Big(\cos(\theta_0)\overline{\psi}_2-\sin(\theta_0)\overline{\psi}_1\Big)
 \mathcal{A} \Big(\cos(\theta_0) {\psi}_2-\sin(\theta_0) {\psi}_1\Big). \label{massbasint}\ee The self-energy for the $\psi_2$ field now includes the diagram depicted in Fig.(\ref{fig:se12}).

\begin{figure}[h]
\begin{center}
\includegraphics[width=12cm,keepaspectratio=true]{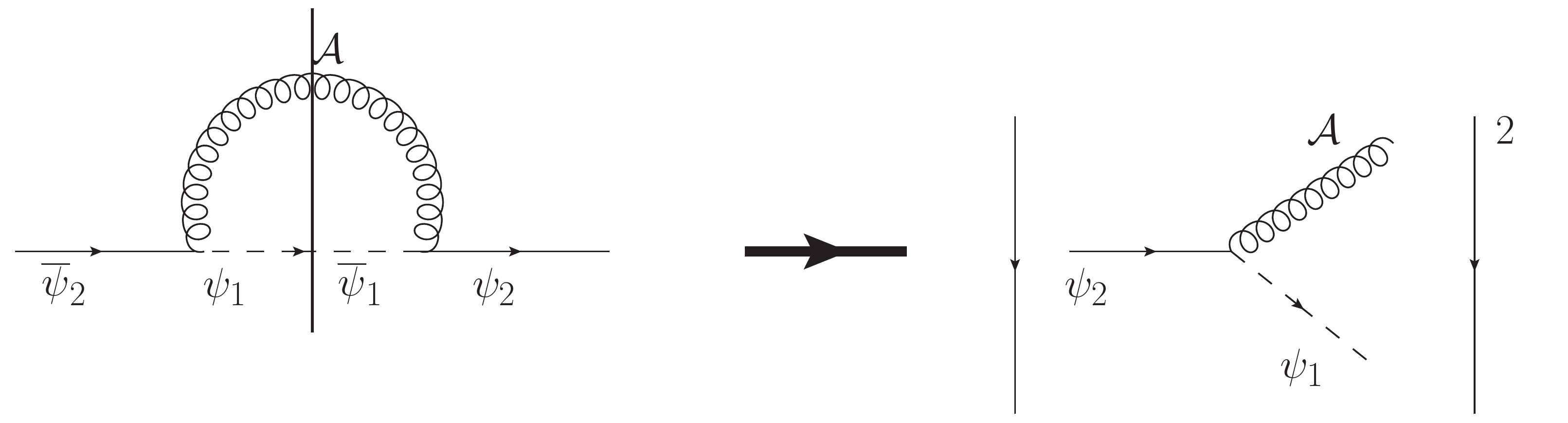}
\caption{Cutkosky cut for the self-energy for mass eigenstate $\psi_2$. The imaginary part
of the self-energy determines the width $\Gamma$ in the pole of the propagator.}
\label{fig:se12cut}
\end{center}
\end{figure}

The cut discontinuity across the intermediate state relates the absorptive (imaginary) part of the self-energy on the mass shell of the external fermion to its decay rate. This relation is depicted in Fig. (\ref{fig:se12cut}).

A standard, straightforward calculation of the decay rate for the process $\psi_2 \rightarrow \mathcal{A}\,\psi_1$, taking
$\mathcal{A},\psi_1$ to be a massless scalar and Dirac fermion, respectively, yields
\be \Gamma_{\psi_2 \rightarrow \mathcal{A}\psi_1} = 2\pi \eta \,M_2\,\Big(\sin(\theta_0)\cos(\theta_0)\Big)^2 \sim \pi \eta \frac{2m^2}{M}, \label{rate}\ee where $\eta$ is given
by Eq. (\ref{eta}) and we have used the approximations $M_2 \approx M;$ $\sin(\theta_0)\approx m/M$ to lowest
order in the see-saw ratio $m/M \ll 1$.

This result coincides to lowest order in $\eta$ and $m/M$ with the non-perturbative imaginary part
of the unparticle-like pole given by Eq. (\ref{gama}).

This simple analysis confirms that
the imaginary part $\Gamma$ in the  propagator of the unparticle-like mode (\ref{alfaminbeta}) describes the
\emph{decay} of the unparticle-like mode into the active-like mode and particles in the hidden
conformal sector. This analysis also validates the interpretation of the width of the unsterile mode (resonance) $\Gamma$ given by (\ref{gama}) as an ``invisible width'' as opposed to the radiative decay width that arises via weak interactions
described  below.

 \section{Cosmological consequences: Radiative decay of the unsterile-like neutrino and the X-ray background}

 Although sterile neutrinos only couple to active neutrinos via an off diagonal mass matrix,
 the diagonalization of this mass matrix results in effective couplings between the sterile-like
 neutrino mass eigenstate and standard model particles, namely active neutrinos and charged
 leptons. Consider the simple case of (canonical) sterile neutrinos coupled to active neutrinos
 via a see-saw mass matrix of the form (\ref{Mmat}) with $F=1$, diagonalized by the usual unitary transformation. In this case,
 \bea \nu_a & = & \cos(\theta_0)\, \nu_1 + \sin(\theta_0)\, \nu_2\,, \label{active} \\
 \nu_s & = & \cos(\theta_0)\, \nu_2 - \sin(\theta_0)\, \nu_1\,, \label{steri}\eea where as usual $\nu_1,\nu_2$ are the
 light (active-like) and heavy (sterile-like) neutrino mass eigenstates, with masses $M_1 \approx m^2/M$; $M_2 \approx M+ m^2/M \gg M_1$, respectively. The mixing angle is determined by Eqs. (\ref{mixangred},\ref{tilC},\ref{tilS}) with
 $F=1$.

\begin{figure}[h]
\begin{center}
\includegraphics[width=12cm,keepaspectratio=true]{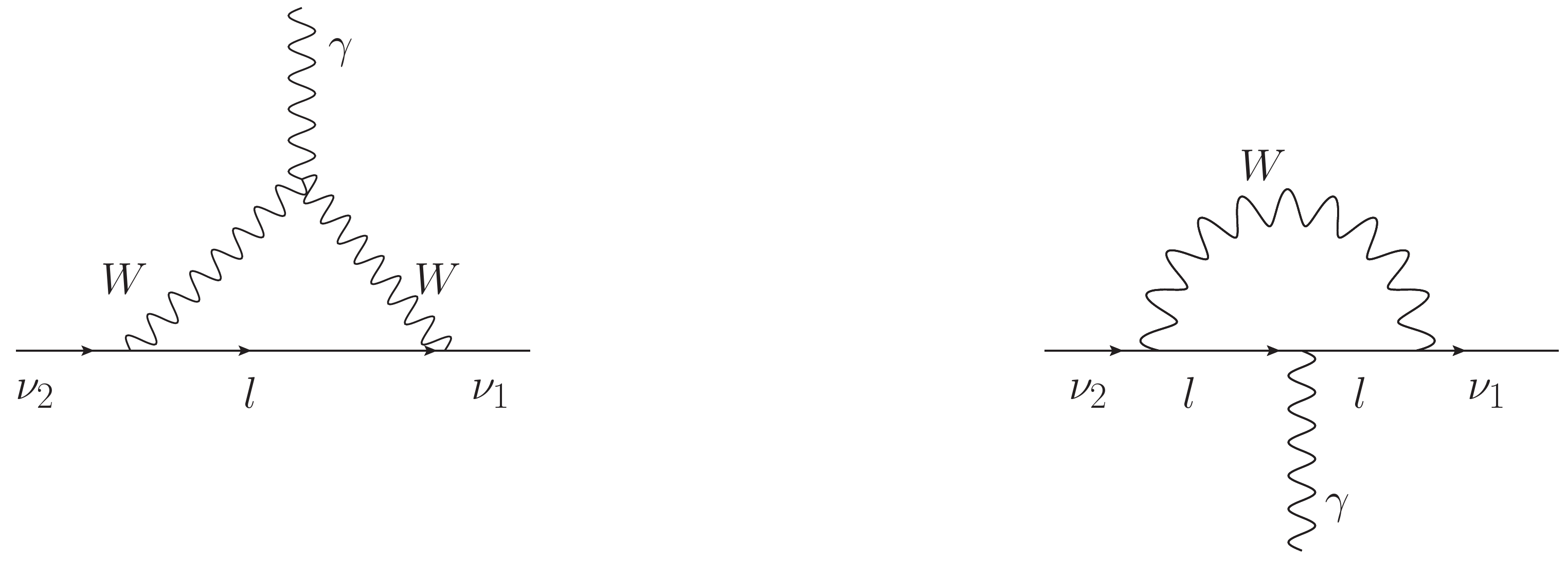}
\caption{Contributions to radiative decay of a sterile-like neutrino in unitary gauge.}
\label{fig:nu2tonu1gamma}
\end{center}
\end{figure}

 The charged current interaction yields an interaction between the sterile-like neutrino and the charged lepton \be \mathcal{L}_{CC}= g\,\overline{\nu}_{aL}\, \not\!{W}\, l_L = g \Big(\cos(\theta_0) \, \overline{\nu}_{1L} + \sin(\theta_0) \, \overline{\nu}_{2L}\Big)\, \not\!{W}\, l_L\,.  \label{CCsm}\ee This interaction vertex leads to the radiative decay of the sterile-like neutrino $\nu_2 \rightarrow \nu_1\,\gamma$ \cite{pal}. The diagrams that describe this process in unitary gauge are
 shown in Fig. (\ref{fig:nu2tonu1gamma}). For $M_2 \gg M_1$, the radiative decay width is given by \cite{pal}
 \be \Gamma_{\nu_2 \rightarrow \nu_1 \gamma} \approx \frac{\alpha_{em}}{2} \Big[\frac{3 G_F}{32\pi^2} \Big]^2 M^5_2~  \Big[\frac{m_l}{M_W}\Big]^4~\sin^2(\theta_0)\,\cos^2(\theta_0)\,, \label{radwidth} \ee where $m_l$ is the
 mass of the charged lepton in the loop in Fig. (\ref{fig:nu2tonu1gamma}).

 If sterile neutrinos are suitable dark matter candidates, with $M_2$ of order of a few keV \cite{kusenko,kusepetra,micha,petra,coldmatter,boysnudm,hector,wu}, the
 radiative decay of the sterile-like neutrino yields a contribution to the X-ray or soft gamma ray background, from
 which stringent bounds on the mass and mixing angle of the sterile-like neutrino are obtained \cite{kusenko,este,kuseobs,micha,hansen,dolgov,aba}.

 The calculation leading to the radiative decay width (\ref{radwidth}) uses the standard assumptions, namely that the propagator of $\nu_2$ is that of a free particle featuring a single particle pole,   that the
  mixing angles are independent of momentum, and that the in and out states are in their (single particle)
  mass shells. All of these assumptions must be revised in view of the results obtained in the
  previous sections. We will proceed to obtain an estimate of the unparticle effects upon the
  radiative decay width by modifying the calculation leading to (\ref{radwidth}) by including the
  following unparticle effects:

  \begin{itemize}
  \item{We will consider the mass eigenstate $\nu_2$ to be described by the propagator (\ref{alfaminbeta}) but neglecting the ``invisible width'' $\Gamma$. This is similar to the situation in calculating
      the decay of a vector boson by considering it to be an asymptotic state with a propagator featuring a single particle
      pole. This assumption restricts the validity of our estimate to $0\leq \eta < 1/3$ since for $\eta > 1/3$, the spectral density of $\nu_2$ is a broad continuum above threshold. Furthermore, since the
      residue at the pole $p^2 \sim M^2_2$ is a finite wave function renormalization $Z_2$, the total transition probability will be multiplied by $Z_2=1/(1-\eta)$. Similarly, the mass eigenstate $\nu_1$ is described by the propagator (\ref{alfaplusbeta}), which does feature an isolated single particle pole at $p^2 = M^2_1$ but with residue $Z_1=1/\big(1+ \eta \Delta^2/2\big)$. Therefore, the transition probability must also
      be multiplied by the wave function renormalization $Z_1$. }

      \item{The mixing angles $\cos(\theta_0) \rightarrow \widetilde{c}(p);$ $\sin(\theta_0) \rightarrow \widetilde{s}(p)$ are given by Eqs. (\ref{candsred},\ref{tilC},\ref{tilS}), which depend on the
          momentum. The radiative decay rate corresponds to setting both the decaying  and the product particles on their  mass
          shells. Therefore,   $ \widetilde{c}(p)$, which is the amplitude of $\nu_1$ in $\nu_a$, must be
          evaluated on the mass shell of the active-like mode, namely $p^2 = M^2_1$, while $\widetilde{s}(p)$, which is the amplitude of $\nu_2$ in $\nu_a$, must be evaluated at the mass
          shell of the unsterile-like mode, $p^2=M^2_2$. Because $M^2_1$ is below threshold, it follows that $ \widetilde{c}(p)$ is real, however, since $M^2_2$ is above threshold,   $ \widetilde{s}(p)$ is complex. Being a probability, the decay
          rate involves the modulus squared of these quantities, namely in the
          expression (\ref{radwidth}), we must replace \bea \cos^2(\theta_0) & \rightarrow &
            \widetilde{c}^2(p)  = \frac{1}{2}\Big[1+  \widetilde{\mathcal{C}}(p)\Big]_{p^2=M^2_1} \,,\label{replaC} \\ \sin^2(\theta_0) & \rightarrow & |\widetilde{s}^2(p)|  = \frac{1}{2}\Big|1-  \widetilde{\mathcal{C}}(p)\Big|_{p^2=M^2_2}\,. \label{replaS}\eea
For $\eta =0$ ($F(p)=1$) and $m\ll M$, it follows that $\cos(\theta_0)\approx 1$, $\sin(\theta_0) \approx m/M$. In the same limit, we find that for $F(p)\neq 1$, $\widetilde{c}(p^2=M^2_1)\approx 1$, $\widetilde{s}(p^2=M^2_2) \approx m/M \sqrt{F(p^2=M^2_2)}$ (see Eqs. (\ref{tilC},\ref{tilS})). The overall change then corresponds to \be \cos^2(\theta_0) \sin^2(\theta_0) \rightarrow \frac{1}{\Big|F(p) \Big| }~ \cos^2(\theta_0) \sin^2(\theta_0)   \,,  \label{totchan}\ee where $p^2 = p^2_2$ is given by (\ref{solplus}).  }

  \end{itemize}

Including all of these modifications we obtain the ratio of the radiative decays for the unsterile, and the (canonical) sterile neutrino as

  \be \frac{\Gamma^U_{\nu_2 \rightarrow \nu_1 \gamma}}{\Gamma_{\nu_2 \rightarrow \nu_1 \gamma}}  =  \frac{\Delta^{\frac{\eta}{1-\eta}}~\Big[ \frac{M^2}{\Lambda^2}\Big]^\eta}{(1-\eta)\,(1+\eta \frac{\Delta^2}{2})}\,. \label{radratio}\ee For a see-saw mass matrix with $m\ll M$, we recognize that
  \be \Delta \approx 2 \sin^2(\theta_0)\,\Big[ \frac{M^2}{\Lambda^2}\Big]^\eta\,, \label{deldef}\ee where $\theta_0$ is the mixing angle for $\eta =0$ (namely the case of canonical sterile neutrinos mixing with active ones). Therefore, we
  can write the ratio (\ref{radratio}) as
  \be \frac{\Gamma^U_{\nu_2 \rightarrow \nu_1 \gamma}}{\Gamma_{\nu_2 \rightarrow \nu_1 \gamma}} \sim
  \frac{ \Big[2 \sin^2(\theta_0) \, \frac{M^2}{\Lambda^2}\Big]^\frac{\eta}{1-\eta}}{(1-\eta)\,(1+\eta \frac{\Delta^2}{2})}\,. \label{radratio2}\ee

  Taking $\sin \theta \ll 1$ and $M < \Lambda$, consistent with a large see-saw and with the unparticle interpretation of the sterile neutrino below a scale $\Lambda$,  respectively, we see that the unparticle nature of the sterile neutrino can lead to a \emph{substantial suppression}
  of the radiative decay rate. Even for $\eta \ll 1$, the current bounds on the mixing angles and   masses for sterile neutrinos from the observations of the X-ray or soft gamma ray background can be weakened considerably. As an example, taking $\sin(\theta_0) \sim 10^{-5}$, $M \sim \mathrm{keV}$ and $\Lambda \sim \mathrm{TeV}$, which are within the range of expectation for physics beyond the standard model, and taking $\eta \sim 0.1$ as an
  example inspired by results in QCD \cite{Neubert:2007kh}\footnote{The value of $\eta$ for QCD obtained in Ref. \cite{Neubert:2007kh}, $\eta \sim 0.5$ is  larger than $1/3$ which limits the validity
  of our result assuming a sharp resonance for $\nu_2$. We have taken $\eta \sim 0.1$ as a representative
  value of the range for  large anomalous dimensions in QCD.}, we find that the ratio (\ref{radratio2}) $\lesssim \mathcal{O}(10^{-3})$.

  This is one of the main cosmological consequences of the unparticle nature of the sterile neutrino.

  \section{Conclusions and further questions}
  In this article we considered the possibility that the $SU(2)$ singlet sterile neutrino might be an unparticle, an interpolating field that describes a multiparticle continuum as a consequence of coupling to a ``hidden'' conformal sector and whose correlation functions feature an anomalous scaling dimension $\eta$. We studied the consequences of its mixing with an active neutrino via a see-saw mass matrix. We focused
  on the simplest setting of one unsterile and one active Dirac neutrino, postponing a more detailed
  study of Majorana neutrinos and several flavors to further study. Our goals here are two-fold:
 \begin{enumerate}
 \item to
  study the consequences of the mixing between the non-canonical unsterile with a canonical active
  neutrino, along with the corresponding dispersion relations and propagating states,
  \item to explore cosmological consequences, in particular the radiative decay width of the unsterile-like neutrino into
  the active-like and a photon, via charged current loops, and to establish how the unparticle nature
  of the sterile neutrino modifies the radiative line-width, which is an important tool to constrain the mass and
  mixing angles of cosmologically relevant sterile neutrinos.
  \end{enumerate}

  We found that the mixing between a non-canonical and a canonical fermion field exhibits several unexpected subtleties. There is no unitary transformation that diagonalizes the full propagator due to the non-canonical nature of the unsterile neutrino. This forces us to make a field redefinition for its complete diagonalization.

  The unsterile-like propagating mode is described by a complex pole above the unparticle threshold for
  $0 \leq \eta < 1/3$, featuring an ``invisible width''. A perturbative analysis and a renormalization group inspired resummation suggest that this width results from the decay of the unsterile-like mode into an
  active-like mode and particles in the conformal sector. As $\eta \rightarrow 1/3^-$, the complex pole
  merges with the unparticle threshold and disappears, while the spectral density of the unsterile neutrino
  features a broad continuum above threshold with a threshold enhancement. The active-like mode corresponds
  to a stable particle, whose propagator features an isolated real pole below the unparticle threshold. This mode ``inherits'' a non-vanishing spectral density above this threshold, even in the \emph{absence}
  of standard model interactions. This novel feature may potentially have relevant consequences since
  the non-vanishing spectral density may open new kinematic channel for standard model processes even to lowest order in weak interactions, a possibility that will be studied in detail elsewhere.

Considering unsterile neutrino as a dark matter candidate, we studied the influence of the unparticle nature of the sterile neutrino on the radiative decay width into an active neutrino and a photon via charged current loops. We find the ratio of decay widths between the
  unparticle case and the canonical case to be \be \frac{\Gamma^U_{\nu_2 \rightarrow \nu_1 \gamma}}{\Gamma_{\nu_2 \rightarrow \nu_1 \gamma}} \approx
  \frac{ \Big[2 \sin^2(\theta_0) \frac{M^2}{\Lambda^2}\Big]^\frac{\eta}{1-\eta}}{(1-\eta)\,(1+\eta \frac{\Delta^2}{2})}\,,\ee where $\theta_0$ is the mixing angle for $\eta =0$, $M $ is approximately
  the mass of the unsterile like neutrino and $\Lambda \gg M$ is the unparticle scale. This ratio
  suggests a substantial suppression of the radiative decay line width for $M \sim \mathrm{keV}$ and
  $\Lambda \sim \mathrm{TeV}$, even for $\eta \lesssim 0.1$. This results in a  weakening of the bounds on the mass and mixing
  angle from the X-ray and soft gamma ray backgrounds.

  Of
  course, a detailed assessment of the suppression of the radiative decay width hinges on the (unknown) values of $\Lambda$ and $\eta$,   which may emerge from the experimental program at LHC in the exploration of physics
  beyond the standard model.
  
To further explore the possibility of an unsterile neutrino as a dark matter candidate, understanding its production process is necessary. Since an unsterile neutrino only interacts directly with the active one, the most effective dark matter production mechanism in this scenario is via unsterile-active neutrino oscillations. It would be interesting to study the implications of our results in Section \ref{mixing} in the dark matter production mechanism along the line of Ref. \cite{dw}. 

\begin{acknowledgments}
D.B. acknowledges support from the U.S. National Science Foundation through Grant No.
 PHY-0553418. R. H. and J. H. are supported by the DOE through Grant No. DE-FG03-91-ER40682.
\end{acknowledgments}

\appendix
\section{An Alternative Explanation for the Non-unitary Transformation   \label{Alt}}
In the helicity basis of Eq. (\ref{PsiRL}), the Lagrangian density in momentum space is given by
\be
{\cal L} =  \sum_h \left( \begin{array}{cc}
                        {\xi_R^h}^{\dagger }(-p) &  {\xi_L^h}^{\dagger}(-p)
                      \end{array} \right) \,  \left( \begin{array}{cc}
                        (p^0 - h |\vp|) \mathds{F} &  \mathds{M}\\
                        \mathds{M} &  (p^0 + h |\vp|) \mathds{F}
                      \end{array} \right) \, \left( \begin{array}{c}
                        \xi_R^h(p) \\  \xi_L^h(p)
                      \end{array} \right).
\ee
Lorentz invariance does not allow us to mix the $(2,1)$-representation of the Lorentz group with the $(1,2)$-representation. Therefore, to diagonalize the action, the allowed transformation is $\xi_{R,L}^h \rightarrow {\cal U}_{R,L} \, \xi_{R,L}^h$, such that all the following matrices:
\be
 {\cal U}_{R} \,\Big[ (p^0 - h |\vp|) \mathds{F}\Big] \, {\cal U}_{R}^{-1},~~ {\cal U}_{L} \,\Big[ (p^0 + h |\vp|) \mathds{F}\Big] \, {\cal U}_{L}^{-1},~~
 {\cal U}_{R} \;\mathds{M} \; {\cal U}_{L}^{-1}~~{\rm and}~~ {\cal U}_{L} \;\mathds{M} \; {\cal U}_{R}^{-1}  \label{umu}
\ee
are all diagonal. Since $\mathds{F}$ is diagonal, and yet not proportional to the identity, the only possible unitary transformations that diagonalize the first two are
\be
{\cal U}_{R,L} =  \left( \begin{array}{cc}
                        1 &  0\\
                       0 &  \pm \, 1
                      \end{array} \right) ~~ {\rm or} ~~ \left( \begin{array}{cc}
                        \pm \, 1 &  0\\
                       0 &  1
                      \end{array} \right).
\ee
However, none of the combinations of these possibilities diagonalize the last two matrices in  (\ref{umu}). Therefore, there is no unitary transformation that diagonalizes the full propagator.

\end{document}